# A New Formula Describing the Scaffold Structure of Spiral Galaxies

Harry I. Ringermacher
General Electric Global Research Center
Schenectady, NY 12309

AND

Lawrence R. Mead
Dept. of Physics and Astronomy
University of Southern Mississippi
Hattiesburg, MS 39406

## ABSTRACT

We describe a new formula capable of quantitatively characterizing the Hubble sequence of spiral galaxies including grand design and barred spirals. Special shapes such as ring galaxies with inward and outward arms are also described by the analytic continuation of the same formula. The formula is $r(\phi) = A/log[B \tan(\phi/2N)]$. This function intrinsically generates a bar in a continuous, fixed relationship relative to an arm of arbitrary winding sweep. $A$ is simply a scale parameter while $B$, together with $N$, determine the spiral pitch. Roughly, greater $N$ results in tighter winding. Greater $B$ results in greater arm sweep and smaller bar/bulge while smaller $B$ fits larger bar/bulge with a sharper bar/arm junction. Thus $B$ controls the "bar/bulge-to-arm" size, while $N$ controls the tightness much like the Hubble scheme. The formula can be recast in a form dependent only on a unique point of turnover angle of pitch – essentially a 1-parameter fit, aside from a scale factor. The recast formula is remarkable and unique in that a single parameter can define a spiral shape with either constant or variable pitch capable of tightly fitting Hubble types from grand design spirals to late type large-barred galaxies. We compare the correlation of our pitch parameter to Hubble type with that of the traditional logarithmic spiral for 21 well-shaped galaxies. The pitch parameter of our formula produces a very tight correlation with ideal Hubble type suggesting it is a good discriminator compared to logarithmic pitch, which shows poor correlation here similar to previous works. Representative examples of fitted galaxies are shown.

*Key Words:* galaxies: spiral – galaxies: structure – galaxies: fundamental parameters

## 1. INTRODUCTION

The logarithmic spiral has been the traditional choice to describe the shape of arms in spiral galaxies. Milne (1946) made perhaps the first attempt to derive these shapes from his own theory, but his theory resulted in spiral orbits for stars. Today most astronomers agree that stellar orbits are essentially circular and that the spiral arms are the result of an evolving pattern, much like a Moire` pattern (Lin & Shu 1964, Shu 1992), or a dynamic modal structure (Bertin et al. 1989a, b; Bertin 1993). That is, the stars define a locus of points at a given time among a family of circular orbits. We shall call this locus an

*isochrone*. The simplest such curve that describes galaxies is the logaritmic spiral and has been used by many (Lin & Shu 1964; Roberts, Roberts & Shu 1975; Kennicutt 1981, Kennicutt & Hodge 1982; Elmegreen & Elmegreen 1987; Ortiz & LéPine 1993; Block & Puerari 1999; Seigar & James 1998a, b, 2002; Seigar et al. 2006; Vallée 2002 ) in their morphological descriptions:

$$r(\phi) = r_0 e^{k\phi} \quad (1)$$

This spiral is usually mathematically characterized by a constant angle of pitch (though k may be a function of r as well) allowing this parameter to be used to describe galaxy shapes. The pitch, *P*, is defined from Binney & Tremaine (1987):

$$\cot(P) = r(\phi) \frac{d\phi}{dr} \quad (1a)$$

For eq. (1), $P = \tan^{-1} k$ is constant. However, it is apparent when attempting fits that galaxy arms often do not have constant pitch. This has also been noted by Kennicutt (1981). This is most evident in strongly barred late type spirals whereas early types and grand designs are essentially constant pitch. In this paper we present a new formula, differing from any in the standard mathematical or astronomical literature, which is capable of describing all spiral shapes, constant pitch or variable, in an elegant way.

## 2. NEW FORMULA

Our formula derives from an examination of equations found in the non-Euclidean geometry of negatively curved spaces. This hyperbolic geometry was first discovered and published by Bolyai (1832) and independently by Lobachevsky. Their work is discussed in Coxeter (1998). The central formula describing multiple parallels measures "the angle of parallelism" ( Coxeter, 1998)  between a given line and "parallel" lines through a given point  not on the line – the violation of Euclid's 5[th] postulate.  The angle of parallelism, known as Lobachevsky's formula is given by $\phi(x) = 2\tan^{-1}(e^{-x})$. The Gudermannian function is closely related and is given by $\phi(x) = 2\tan^{-1}(e^{x})$. The latter function directly relates circular to hyperbolic functions. We have found a new function closely related to the above that describes the shapes of spiral galaxies remarkably well. This formula is given in radial form, where $r^{-1}$ replaces $x$ in the Gudermannian and scaling degrees of freedom are added:

$$r(\phi) = \frac{A}{\log\left(B \tan \frac{\phi}{2N}\right)}. \quad (2)$$

This function intrinsically generates a bar in a continuous, fixed relationship relative to an arm of arbitrary winding sweep. Though in some instances, observations show gaps between the bar and arms (e.g., Seigar & James 1998b), nevertheless, arms begin where bars end so that a continuous bar-arm formula serves as a galactic fiducial for fitting. This is particularly evident in NGC 1365 of our galaxy selection and will be described later. *A* is simply a scale parameter for the entire structure while *B*, together with a new parameter *N*, determine the spiral pitch. The "winding number", *N*, need not be an integer. Unlike the logarithmic spiral, this spiral does not have constant pitch but has precisely the pitch variation found in galaxies. The use of this formula assumes that all galaxies have "bars" albeit hidden within a bulge consistent with recent findings. Roughly, greater $N$ results in tighter winding. Greater $B$ results in greater arm sweep and smaller bar/bulge while smaller $B$ fits larger bar/bulge with a sharper bar/arm junction. Thus $B$ controls the "bulge-to-arm" size, while $N$ controls the tightness much like the Hubble scheme. Figure (1) shows several examples of these spirals. We divide the examples according to $N$-value. The opposing arm is added by symmetry. Scale plays an important role in that the interior of the same spiral, when expanded could fit a barred galaxy as well as a grand design. This is demonstrated in Fig. 1a where the scale factor, $A$, has been increased a factor of 6 over the remaining examples ($A = 1$). The examples range from barred spirals to grand designs and large arm sweeps.

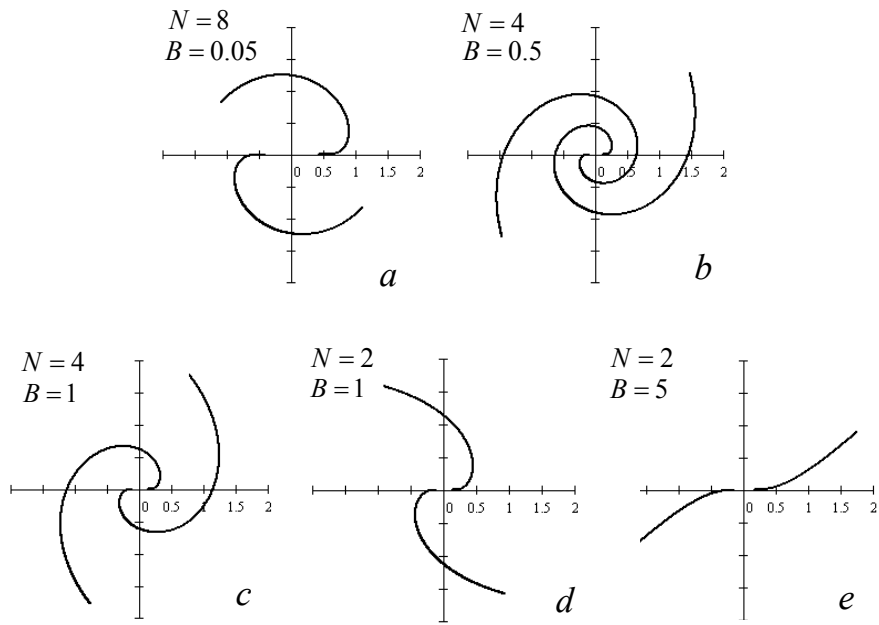

Fig.1.   Examples of Eq. (2) for varying $N$ and $B$

## 3.  GALAXY FITS

*Down-Projection and Up-Projection*

Galaxy shapes in the sky are projections with respect to a North-South, East-West coordinate system which we simply define as oriented along Y and X axes respectively on a graph facing us. Two angles, namely position angle (PA) and inclination angle (I), are necessary to down-project a shape from a "sky-plane" to a "graph-plane". By the previous definition, the two planes are actually one and the same. The end result of a 2-angle down-projection, PA followed by I, is a correct but oriented graph shape at a third angle, γ. In this case, the third angle is the orientation, γ, in the plane with respect to "Y" or N-S. We recognize the shape in any direction so it is not important.

This procedure is, however, not arbitrarily reversible. If one creates a theoretic shape function to compare to an observed galaxy and simply starts with the major axis aligned along "Y", then up-projects using the known I followed by PA (reversing the order and sign of angle), the shape would, in general, be incorrect and we would require a third Euler rotation, γ. Alternatively, we could apply the "final orientation", γ, determined from down-projection as the first rotation about Z, and then apply I followed by PA and find the correct sky-shape. Equivalently, one could replace $\phi$ in formula (2) by $\phi - \gamma$ and achieve the same effect. It is clear from either view that the third angle is necessary for up-projection otherwise a serious shape error could result. The necessity for a third angle is most obvious in cases where a galaxy shape is not equiaxial in its plane. There are then 2 unique axes in the sky plane, the major axis as viewed and the intrinsic long axis and thus the need for a third angle to reconcile them. Circularly symmetric tightly wound spirals and face-on galaxies do not require a third angle, but many other structures, as will be demonstrated, do.

*Galaxy Fits*

We have fitted many galaxies with formula (2). Below we present fits to a variety of spiral galaxy shapes, some of which are difficult to describe with any other formula. The polar isochrone can be rotated through three Euler angles ($\alpha, \beta, \gamma$) about the (Z,Y, Z) axes to best fit the observed galaxy. In principle, the three Euler angles define an arbitrary rotation in a 3-space uniquely. Here we define the three rotations as follows: the first rotation, $\alpha$, CCW about the Z-axis out of the graph plane; the second rotation, $\beta$, CW about the rotated Y-axis in the graph plane; and the third rotation, $\gamma$, CCW about the rotated Z-axis. The angle $\alpha$ is the position angle and $\beta$ is the inclination angle when $\gamma$ is not needed and the image is correctly sky-oriented. We shall call the third angle, $\gamma$, the "twist". The more circular a galaxy shape is or the more face-on it is, the less the need for twist. The three angles fit rather tightly. Typically a few degrees variation shows significant differences in the global fit. Figure (2) shows a best eye-fit of formula (2) and the log-spiral (1) to NGC 1365, a classic barred spiral, traditionally classified SBbc. Pre-rotated graphs are seen in the lower left. Cloned galaxies are shown in the upper right for clarity. It is seen that a log-spiral with an 18° pitch from Kennicut (1981) cannot fit over the full range of the arms. In this case a good match was chosen near the arm-bar junction. A good match could have been chosen along the distant arms or an average match could have been chosen. What is clear is that this galaxy has a seriously variable pitch. Traditionally, an "average" pitch is chosen and is obtained by a variety of methods. Unlike our "eye-fit" of 2π or greater, these average matches are taken over varying radial intervals and do not, in general, sample all the available range. For example, although both Kennicut and Seigar use averaging, Kennicut (1981) finds the average pitch angle for NGC 1365 to be 18° while Seigar (2006) finds 35°. Clearly, Seigar's analysis favored an interior (near the bar-arm junction) average while Kennicutt's favored an exterior (outer arms) average. We found that the outer arm pitch approached a 10° limit while the innermost pitch was far greater than the Seigar value. It is no wonder that a common value cannot be agreed upon. How good the agreement is depends strongly on the precise point chosen for the pitch origin. Both the method of fitting (here, a global fit) and the presence of "twist" will affect the pitch origin. This is demonstrated in Fig.3, where NGC 1365 is fitted with zero twist. The bar-arm junction is severely mismatched thus dislocating the pitch origin. An average pitch for this fit would favor an "exterior" value since the pitch origin is well away from the junction. Figure 4 shows a fit of both equations to M51. Both are excellent fits indicating this grand design spiral is close to constant pitch. Figure (5) shows a fit to NGC 1097, classified SBb. This is essentially the same shape as NGC 1365 with fitting parameters ($N=16, B=0.4$), but differing arm length and position. The log-spiral (8° pitch) is very good for most of the exterior arm but fails along the interior due to varying pitch. The Kennicutt pitch is 17°. Figure (6) shows a fit to NGC 1300, also SBb, which again has parameters ($N=16, B=0.4$) suggesting that large barred galaxies may have a universal shape. NGC 1300 shows some deviation in the upper arm, but the formula assumes perfect, symmetric arms. Deviations are not expected to be fit for any number of causes. Note that the formula acts as a "scaffold" description and will not create the detailed inner bar structure but rather a continuous bar replacing it.

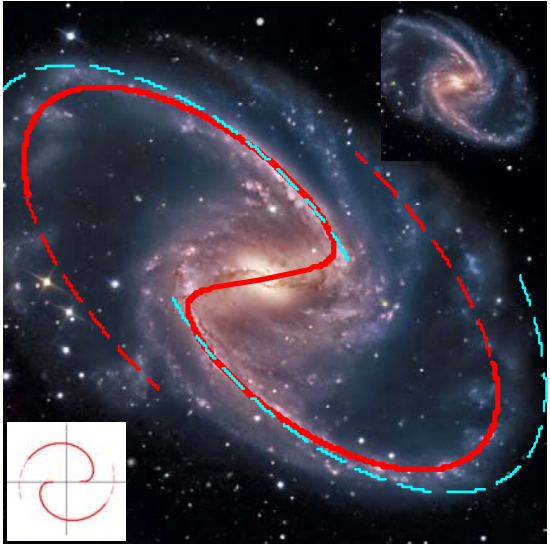

Fig. 2. NGC 1365: best fit isochrone(red) from eq. (2). (2). N=16, B=0.4, Euler angles (47,62,18). Log- spiral (dashed-cyan): 18° pitch. Credit:NOAO/AURA/NSF

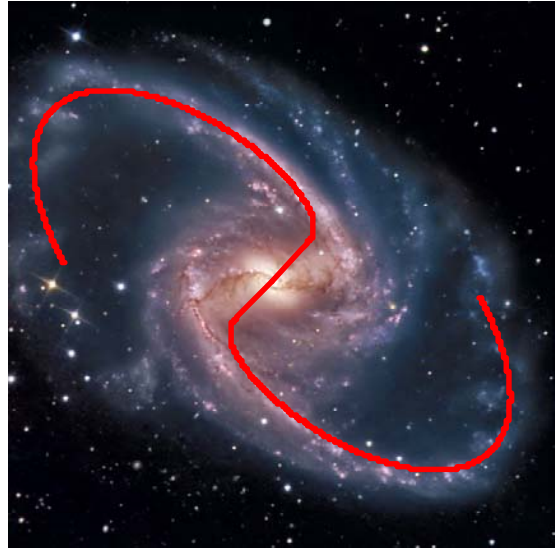

Fig. 3. NGC 1365: best fit isochrone from eq. N=16, B=0.4, Euler angles (47,62,0).

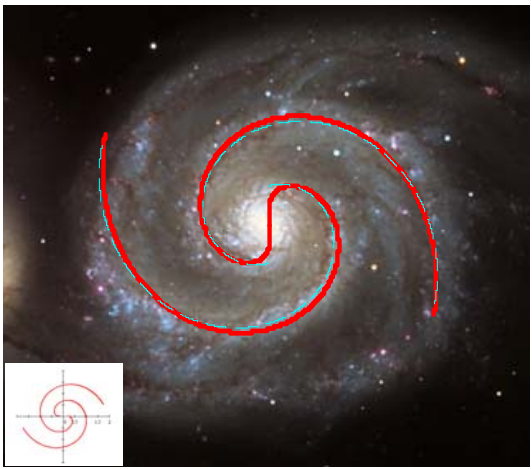

Fig. 4. M51: best fit isochrone from eq. (2). N=4, B=0.63, Euler angles (90,0,0). Log-spiral (dashed-cyan): 17° pitch.

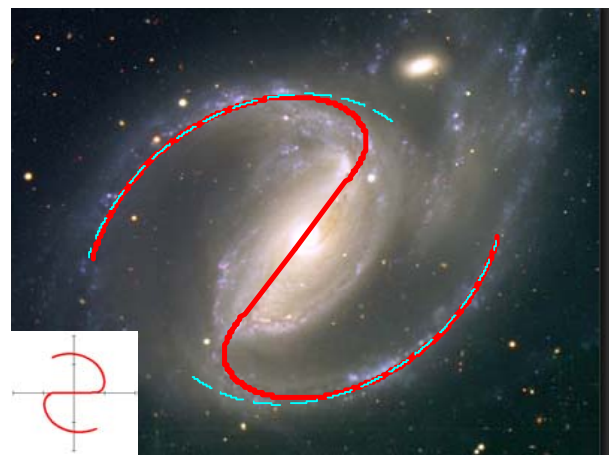

Fig.5. NGC 1097: best fit isochrone(red) from eq. (2). N=4, B=0.08, Euler angles (52,37,23). Log-spiral (dashed-cyan): 8° pitch.

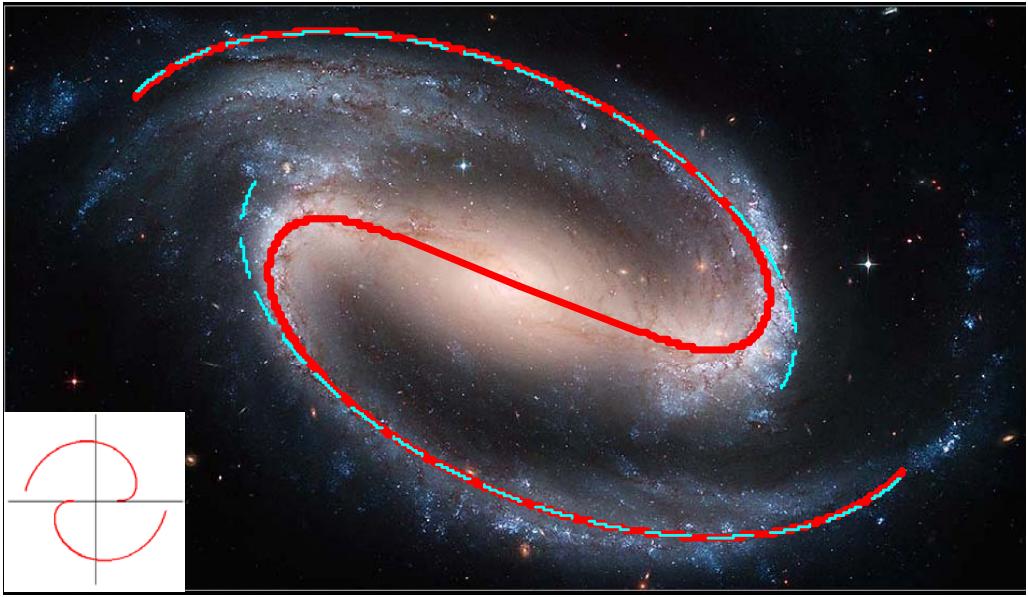

Fig. 6. NGC 1300: best fit isochrone from eq. (2). N=4, B=0.08, Euler angles (65,55,79). Log-spiral(dashed- cyan): 9° pitch .

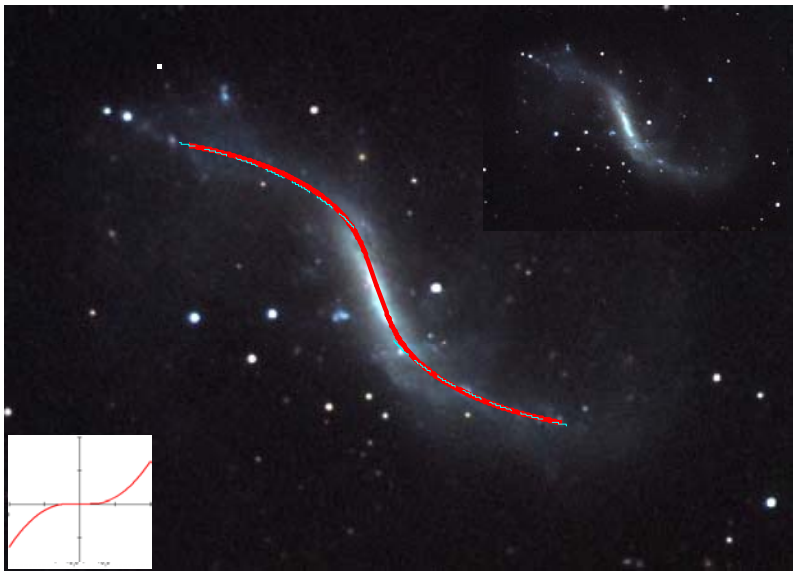

Fig. 7. NGC 4731: best fit isochrone from eq. (2).N=2,B=3, Euler angles (110,0,0). Log-spiral(dashed- cyan). 67° pitch.

Figure 7 shows a best fit to NGC 4731 for ($N = 2, B = 3$). This galaxy can be equally well fitted by the logarithmic spiral, equation (1), for the large pitch factor, k = 2.3, corresponding to 67°.

Ring galaxies are a special class that cannot be described by eq. (1). However an analytic continuation of formula (2), where tangent is replaced by hyperbolic tangent, is capable of describing ring galaxies with spiral structure. The analytic continuation is obtained by setting $B \equiv 1/\tanh(\phi_0/2N)$ and replacing $(\phi,\phi_0) \to (i\phi, i\phi_0)$ to yield:

$$r(\phi) = \frac{A}{\log\left(B \tanh \frac{\phi}{2N}\right)} \tag{3}$$

Figure 8 shows NGC 4622, classified SAb, fitted with formula (3). This formula produces rings with either ingoing or outgoing spirals. A log-spiral with zero pitch would generate a ring – but no arms. Unlike a log-spiral, this formula generates both. The parameters used were; (outgoing: $N = 7$, $B = 1.75$; ingoing: $N = 4$, $B = 0.4$). In this case several rings were matched and overlaid to fit this unusual galaxy structure subject to the constraint that all arms emanate from a single ring. The spiral structure here is particularly sharp and well fitted by the formula. The outward arms are leading while the inward arms (blue) are trailing in this "reverse" galaxy.

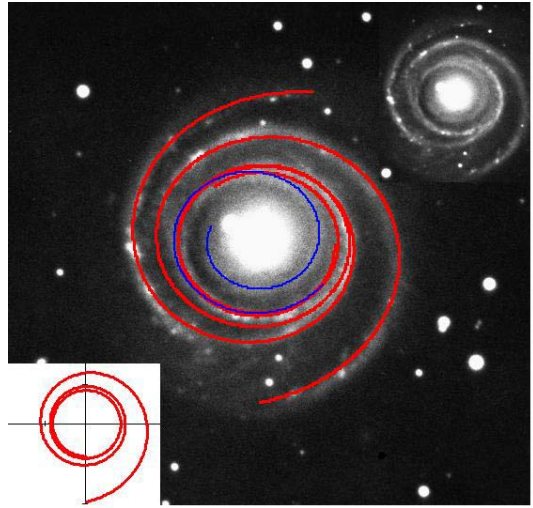

Fig. 8. NGC 4622: best fit isochrone from hyperbolic eq. (3). Euler angles (0,23,0).

## 4. FORMULA USING ANGLE OF PITCH

Astronomers generally use an angle of pitch to describe the shape of spirals. Formula (2) can be renormalized to accommodate a referenced angle of pitch replacing $B$. The angle of pitch is defined as the angle between the tangent to the curve at a given point $(r,\phi)$ and the tangent to a circle of radius $r$ through the point. The renormalization of (2) is described in Appendix A. The result is a unique formula, referenced only to the angle $\Phi$, the angle of pitch at "turnover" (see Appendix A):

$$r(\phi) = \frac{R_\Phi}{1 - \Phi \tan(\Phi) \log\left(\frac{\phi}{\Phi}\right)} \tag{4}$$

We do not yet have an equivalent renormalization of formula (3). For a unit bar radius, the single parameter, $\Phi$, determines the shape of spirals with *nearly constant or variable* pitch. Figure 9 shows examples of the use of (4) for Hubble classes Sa, Sb and Sc with $\Phi$ varying from 0.4 to 1.0 (9a through 9e). For larger $\Phi$, (9f), the arm no longer turns over. An example of this shape is NGC 4731 (Figure 7).

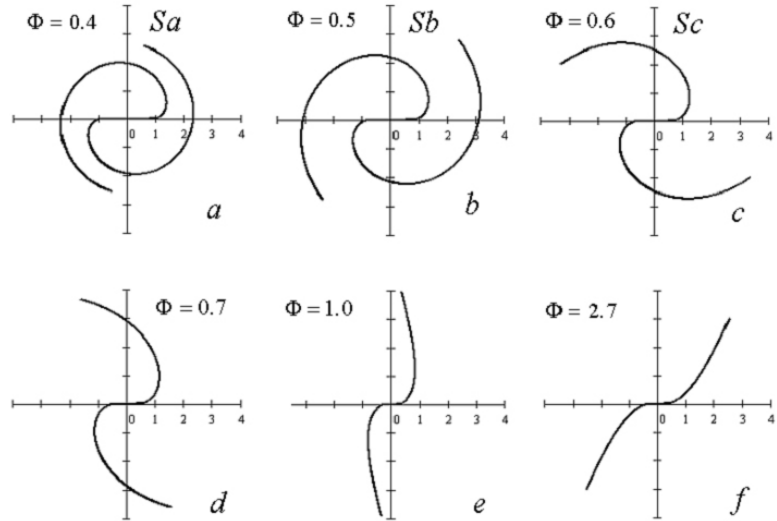

Fig. 9. Examples of Eq. (4) for various "turnover" pitch angles, $\Phi$, with Hubble classes indicated.

## 5. CORRELATION OF "TURNOVER" PITCH WITH HUBBLE TYPE

Since the Hubble scheme is a simple morphological classification organizing galaxy shapes in terms of their arm sweep, bulge size and relationships between the two, one might expect a strong correlation between the arm pitch angle and the classification parameter. Understanding that such a scheme is qualitative, depending strongly on the observer, however, does not explain why, to date, there is essentially no correlation (Seigar et al. 2006). Kennicutt (1981) found a very weak correlation at best. It is therefore of interest to examine the relationship of our parameter, $\Phi$, the angle of pitch at turnover, to Hubble type. We have selected 21 well-defined galaxy shapes from the tables of Seigar (1998, 2006), Kennicutt (1981), and Rubin et al. (1985) and evaluated $\Phi$ for each from formula (4) by a best-eye global fit as exemplified in figures 2 through 8 (Table 1). These fits are very tight with variations of only a few degrees causing significant deviations about each rotation. This is an iterative process. First a rough shape is chosen based on the standard classification using (N, B) parameters of equation 2. Then position angle is easily set while a first estimate of inclination is taken from the literature and fine-tuned. In most instances this is insufficient for a good fit and twist is necessary. N and B are then fine-tuned for best fit. The $\Phi$ parameter can be calculated numerically from equation A8 for a given (N, B) pair. Equation (2) can be degenerate in that two (N, B) pairs can result in essentially the same fit. For example, NGC 1365 (Figure 2) was fit by (16, 0.4) but can equally well be fit by (4, 0.08). However, both of these pairs result in the same $\Phi$ parameter, 0.42 radians, within 2%. Thus the $\Phi$ parameter is a unique shape discriminator. Alternatively, knowing the Φ behavior of equation (4), one can use it directly to fit the shape. The Φ parameters for the 21 galaxies are shown in Table 1. We also best eye-fit a logarithmic spiral, equation (1), to the galaxies. Although these matched our fit often, there were, on average, interior and exterior deviations. The constant pitch *P*, in degrees, for each fit is also shown in table 1, column 5.  The NGC 1365 pitch variation was so severe that Kennicutt's "average" of 18° was chosen as a compromise between our fit at 10° and Seigar's "average" at 35°. The de Vaucouleurs (1959) numerical stages are displayed in column 2 ( a =1, b = 3, c = 5) corresponding to the Hubble type of column 6 or corrected from column 7, a relabeling found to produce consistency among all the a,b and c categories. The basis for this relabeling was first suggested by

Table 1
Hubble types and arm pitch of 21 spiral galaxies

| Galaxy | Type | Φ rad | Φ deg | P deg | Type | Relabels |
|---|---|---|---|---|---|---|
| M 51 | 5 | 0.52 | 29.79 | 16.7 | SAc | |
| M 81 | 5 | 0.512 | 29.34 | 11.3 | SAb | SAc |
| NGC 4321 | 5 | 0.535 | 30.65 | 14.6 | SAc | |
| NGC 2997 | 5 | 0.543 | 31.11 | 15.6 | SAc | |
| M 74 | 5 | 0.526 | 30.14 | 15.6 | SAc | |
| NGC 3198 | 5 | 0.537 | 30.77 | 15.1 | SAc | |
| UGC 4643 | 4 | 0.453 | 25.96 | 8.0 | SAbc | |
| NGC 5364 | 4 | 0.474 | 27.16 | 10.2 | SAc | SAbc |
| NGC 1097 | 3 | 0.415 | 23.78 | 8.0 | SBb | |
| NGC 1300 | 3 | 0.415 | 23.78 | 9.1 | SBb | |
| NGC 1365 | 3 | 0.423 | 24.24 | 18.0 | SBbc | SBb |
| NGC 7096 | 3 | 0.398 | 22.80 | 8.5 | SAa | SAb |
| NGC 1357 | 3 | 0.395 | 22.63 | 5.7 | SAa | SAb |
| NGC 4593 | 3 | 0.427 | 24.47 | 9.1 | SBb | |
| NGC 1417 | 3 | 0.42 | 24.06 | 9.1 | SABb | |
| NGC 5754 | 3 | 0.425 | 24.35 | 9.1 | SAb | |
| NGC 266 | 3 | 0.397 | 22.75 | 11.3 | SBab | SBb |
| NGC 3281 | 3 | 0.415 | 23.78 | 6.3 | SAab | SAb |
| NGC 1398 | 2 | 0.383 | 21.94 | 4.6 | SBab | |
| NGC 3504 | 1 | 0.337 | 19.31 | 4.6 | SBb | SBa |
| NGC 4622 | 1 | 0.362 | 20.74 | 4.0 | SAb | SAa |
| NGC 2273 | 1 | 0.346 | 19.82 | 4.0 | SAa | |
| NGC 4731 | 4 | 2.569 | 147.19 | 66.5 | SBc/P | |

Kennicutt (1981), where he indicated that certain morphological features could generate an inconsistency between arm pitch angle and expected Hubble type. As seen, column 2 includes the relabeling corrections.  Relabeling resulted in shifting 5 galaxies by one category (e.g. b → c) and 4 galaxies within one category (e.g. bc → b). Figure 10 shows the correlation of Φ (degrees) with Hubble type using the corrected type, column 2. The goodness of fit to the straight line,

$$\Phi = 2.69 \cdot (deVauc\ Hubble\ stage) + 16.22\ ,$$

is 94%. If this re-assignment is not made, our own parameter also shows only weak correlation (Figure 11) with 62% goodness of fit.

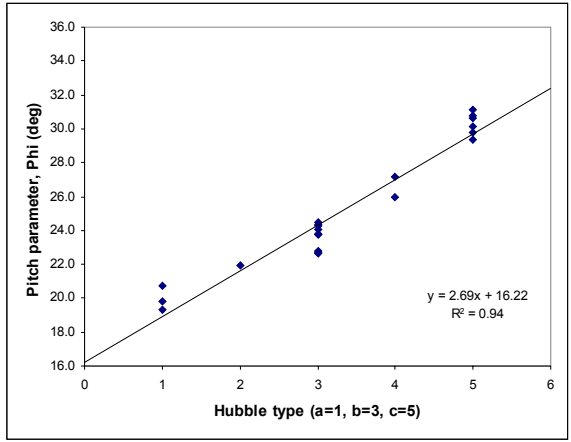 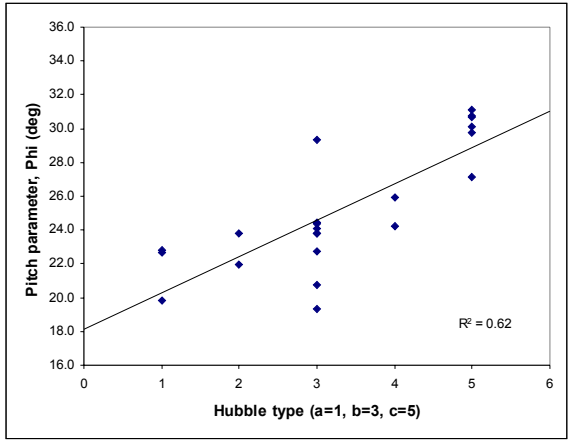

Fig. 10. Pitch parameter Φ vs. relabeled Hubble type    Fig. 11. Pitch parameter Φ vs. "as is" Hubble type

Even more interesting is a plot of the constant log-spiral pitch, *P*, versus Hubble type (Figure 12). These are painstakingly fitted values and in most cases closely matched our fit from equation (4) but with averaged deviations. *Nevertheless, even using the relabeled Hubble type values, the log-spiral pitch shows only a weak correlation of 64% with type, but similar slope to ours*. When *P* is plotted against the standard type values (Figure 13), the correlation is worse at 58% but not much worse. This is strong evidence that the log-spiral itself is inadequate to describe spiral galaxy shapes and its failure is only compounded by misclassification. To prove that our relabeling shifts result in a self-consistent morphological description, we overlay the graph-plane 21 shapes in the relabeled categories a, b and c. The overlays for each type, using our equation, are very nearly identical and clearly distinguishable thus justifying the class relabeling. The overlays were made after normalizing the graph-plane shapes so that an arm intercepts the x-axis at unit length after a CW $\pi$ arm rotation.

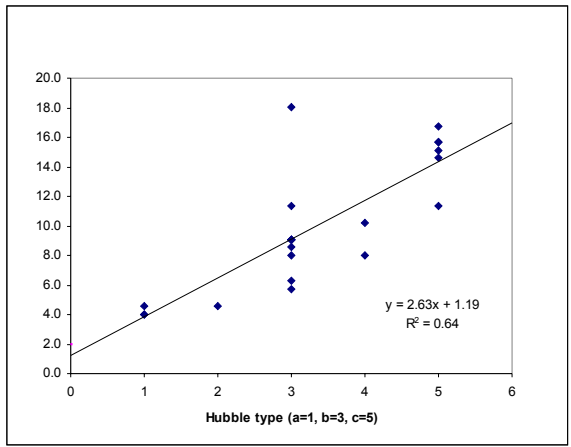 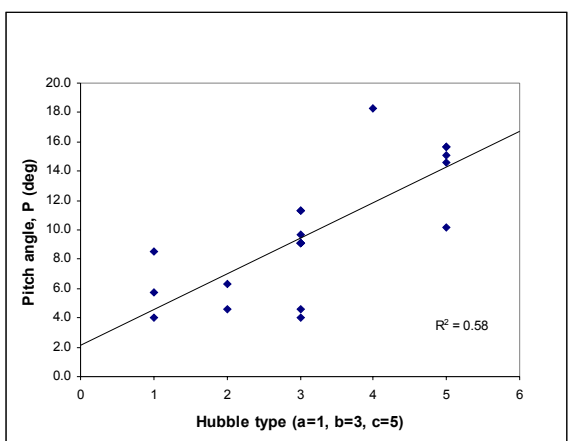

Fig. 12. Log-spiral pitch P vs. relabeled Hubble type    Fig. 13. Log-spiral pitch P vs. "as is" Hubble type

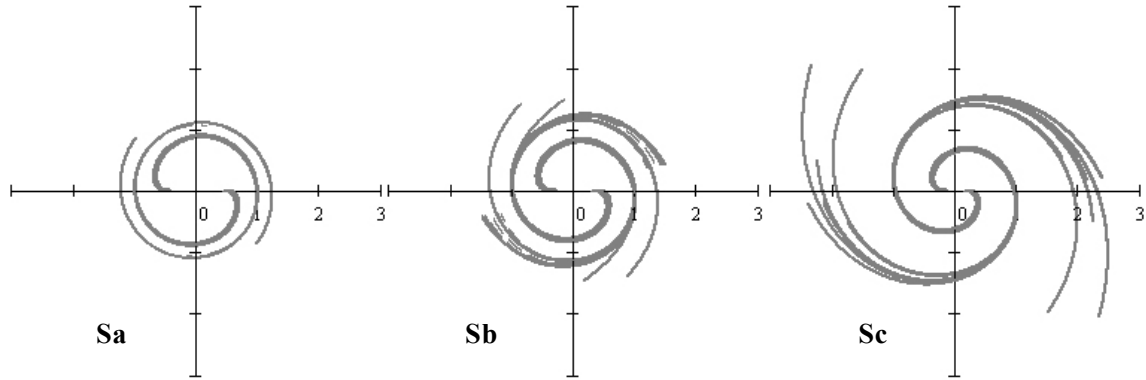

          **Sa**                           **Sb**                         **Sc**

Fig. 14. Overlays of 21 galaxy de-projections showing consistency of Sa, Sib and Sc shapes, based on arm normalization

This procedure gives a correct perspective on the bar (bulge) arm relationship as can be seen progressing from Sa with a large bar/tight-arms to Sc with a small bar/sweeping-arms. All but two galaxies fall precisely in these three shapes – exactly so for $\pi$ rotation and nearly so up to $2\pi$ rotation in types a and c. The two mid-class galaxies literally do not fit these three and fall between.

## 6. DISCUSSION

    We have shown that the constant pitch logarithmic spiral is an inadequate discriminator of Hubble type for spiral galaxies, which basically explains why poor correlations with type are the norm. The pitch, of course, can be made variable, but that would introduce additional parameters dependent on each fitted galaxy. We present an new, elegant, *single parameter* formula, closely related to non-Euclidean geometric functions, with an intrinsically varying pitch that describes all the Hubble classifications faithfully. This function has a natural, correctly proportioned, bar continuously transitioning to the arms that serves as a shape fitting fiducial permitting the extraction of a tightly discriminating pitch parameter. Its analytic continuation naturally describes spiral ring structures with ingoing or outgoing arms – something not achievable from a logarithmic spiral. The fits of these new functions to galaxies are remarkable. The correlation of the new pitch parameter to ideal Hubble type is excellent – only when a number of galaxies are reclassified for self-consistency. Without reclassification, no strong correlation of arm pitch to Hubble type can ever be expected for any formula.

    With the current interest in morphological evolution, it may be desirable to have a reliable quantitative classifier of galaxies. We have presented two formulas. The simpler one, (2), is essentially a two-parameter fit. The parameters can, however, be degenerate. Formula (4) is a renormalized version of (2) that is self-referenced to the angle of pitch at the spiral "turn-over" point and is unique for every shape. Formula (4) reproduces,

precisely, all the shapes of (2) for appropriate choices of the parameter $\Phi$, including sharp arm-to-bar junctions, at small values, suitable for some barred spirals. We have made an initial attempt at parametric classification (see Figure 9). Spirals with pitch parameter less than 0.40 radians might be classified as Sa while spirals with pitch parameter between 0.4 and 0.5 radians might be classified as Sb. The class Sc, with pitch parameter greater than about 0.5 radians, has a broad range of sweep. We have not included class d through m because these are even more qualitative. These are preliminary judgements. In order to use these formulas properly, one must understand their parametric behavior, angle range and applicability for many more well-shaped galaxies. A practical application of this formula, for example, to automated classification, is possible – since the global fitting procedure is well defined - but is beyond the scope of this paper.

## APPENDIX A

The pitch angle, $P$, from (1a) applied to the isochrone (2) is given by:
$$U \equiv cot(P) = r(\phi)\frac{d\phi}{dr} = -N\,sin(\phi/N)\,log[B\,tan(\phi/2N)] \quad (A1)$$
It can be shown that the unit tangent vector to the isochrone is:

$$\hat{T} = \frac{(cos\,\phi - U\,sin\,\phi)\hat{i} + (sin\,\phi + U\,cos\,\phi)\hat{j}}{\sqrt{1+U^2}} \quad (A2)$$

Now, we wish to reference the isochrone to a particular point, $(R_\Phi, \Phi)$, on the curve. We can therefore write:
$$R_\Phi = \frac{A_1}{log[B\,tan(\Phi/2N)]} = \frac{A_1}{log[B] + log[tan(\Phi/2N)]} \quad (A3)$$
Solving for $log[B]$:
$$log[B] = \frac{A_1}{R_\Phi} - log[tan(\Phi/2N)] \quad (A4)$$
Thus, referenced to a particular point $(R_\Phi, \Phi)$, the isochrone, (2), may be written:

$$r(\phi) = \frac{R_\Phi}{1 + \frac{R_\Phi}{A_1}\,log\left(\frac{tan(\phi/2N)}{tan(\Phi/2N)}\right)} \quad (A5)$$

*Choice of* $(R_\Phi, \Phi)$

A convenient choice might be an angle that approximates $R_\Phi$ as the "bar radius". One such unique angle is at the "turn-over" point of the isochrone where the tangent vector points along $\hat{j}$. From (A2), this condition is:
$$tan\,\Phi = 1/U \quad (A6)$$
From (A6) and (1a), we find

$$\Phi = P. \tag{A7}$$

That is, the angle $\phi$ at turn-over is precisely the angle of pitch at that point. So this is indeed a unique point. From (A6), using the definition of $P$ from (1a), one finds:

$$log[B\,tan(\Phi/2N)] = \frac{-1}{N\,sin(\Phi/N)\,tan(\Phi)} \tag{A8}$$

From (A3) and (A8) we also have:

$$\frac{R_\Phi}{A_1} = -N\,sin(\Phi/N)\,tan(\Phi) \tag{A9}$$

We use this relation to fully re-normalize the isochrone with respect to $\{R_\Phi, \Phi, N\}$;

$$r(\phi) = \frac{R_\Phi}{1 - N\,sin(\Phi/N)\,tan(\Phi)\,log\left(\frac{tan(\phi/2N)}{tan(\Phi/2N)}\right)}, \tag{A10}$$

where: $\{R_\Phi, \Phi, N\} \equiv$ {bar radius, angle of pitch at turnover, winding number}. For grand design spirals, replace "bar radius" by "bulge radius". Note that (A8) can be used to relate the $(\Phi, N)$ formula (A10) to the original $(B, N)$ formula (2) by solving numerically for $\Phi$, given $B$. Typical $N$ factors range from 2 to 16.

For $N \geq 2$, we note that relation (A10) is essentially independent of $N$ with excellent fits to all the previous images and plots. This can be seen by using a small angle approximation for functions containing $N$. The pitch at this point is typically around 30°. Significant errors accumulate in the fits for $N < 2$.
That is to say, the following formula is a good approximation for $N \geq 2$ - nearly all cases:

$$r(\phi) = \frac{R_\Phi}{1 - \Phi\,tan(\Phi)\,log\left(\frac{\phi}{\Phi}\right)} \tag{A11}$$

## REFERENCES

Binney, J. & Tremaine, S., 1987, Galactic Dynamics, Princeton U. Press

Bertin, G., Lin, C.C., Lowe S.A., Thurstans R.P.,1989a, ApJ, 338, 78

Bertin, G., Lin, C.C., Lowe S.A., Thurstans R.P.,1989b, ApJ, 338, 104

Bertin, G., 1993, ASP, 105, 640

Block, D.L., Puerari, I., 1999, A&A, 342, 627

Bolyai, János, 1832, Appendix: Scientiam spatii absolute veram exhibens: a veritate aut falsitate Axiomatis XI Eucldei…(Ref: Gray, Jeremy, J., 2004, János Bolyai, Non-Euclidean Geometry, and the Nature of Space, Burndy Library Publications)

Coxeter, H.S.M., 1998, Non-Euclidean Geometry, 6th Ed., Mathematical Association of America

De Vaucouleurs, G., 1959, Handbuch der Physik, 53, 275
Elmegreen, D.M. & Elmegreen, B.G., 1987, ApJ, 314, 3
Kennicutt, R.C., 1981, AJ, 86, 1847
Kennicutt, R. & Hodge, P., 1982, ApJ, 253, 101
Lin, C.C. & Shu, F.H., 1964, ApJ, 140, 646
Milne, E.A., 1946, MNRAS, 106, 180
Ortiz, R. & LéPine, J.R.D., 1993, A&A, 279, 90
Roberts, W.W., Roberts, M.S., Shu, F.H., 1975, ApJ, 196, 381
Rubin, V., Burstein, D., Ford, K., Thonnard, N., 1985, ApJ, 289, 81
Shu, Frank H., 1992, Gas Dynamics, University Science Books
Seigar, M.S. & James, P.A., 1998a, MNRAS, 299, 672
Seigar, M.S. & James, P.A., 1998b, MNRAS, 299, 685
Seigar, M.S. & James, P.A., 2002, MNRAS, 337, 1113
Seigar, M.S., Bullock, J.S., Barth, A.J., Ho, L.C., 2006, AJ, 645, 1012
Vallée, J.P., 2002, ApJ, 566, 261